\documentclass[draft,onecollarge]{svjour2}

\usepackage{amsfonts}
\usepackage{amsmath}
\usepackage{amssymb}
\usepackage{mathptmx}
\usepackage{pstricks}

\spnewtheorem{thm}{Theorem}{\upshape\bfseries}{\itshape}
\spnewtheorem{rmk}{Remark}{\upshape\bfseries}{\upshape}
\spnewtheorem{dfn}{Definition}{\upshape\bfseries}{\upshape} 
\usepackage{tikz}
\usetikzlibrary{%
  calc,arrows,matrix,positioning,snakes%
}

        \tikzset{%
         line cap=round,%
         ampersand replacement=\&,%
         injective^/.style={right hook->},%
         injective_/.style={left hook->},%
         surjective/.style={->>},
         bijective/.style={right hook->>},
         claim/.style={dotted,line cap=round},%
         commutative-diagram/.style={%
         matrix of nodes,column sep=.9cm,row sep=1cm,%
         every node/.style={anchor=base,text height=1.5ex,text depth=.25ex}},%
         commutative-diagram-arrows/.style={%
         every node/.style={midway,font=\scriptsize,%
         text height=1.5ex,text depth=.25ex,inner sep=2pt}}%
        }

\let\bb\mathbb
\let\vek\vec
\let\defemph\emph
\let\inrightarrow\hookrightarrow

\makeatletter
\renewcommand*\env@matrix[1][c]{\hskip -\arraycolsep
  \let\@ifnextchar\new@ifnextchar
  \array{*\c@MaxMatrixCols #1}}
\makeatother

\def\xfrac#1#2{\mathord{\mathchoice%
   {\raise1pt\hbox{$\displaystyle #1$}/%
     \lower1pt\hbox{$\displaystyle #2$}}
   {\raise1pt\hbox{$\textstyle #1$}/%
     \lower1pt\hbox{$\textstyle #2$}}
   {\raise.7pt\hbox{$\scriptstyle #1$}/%
     \lower.7pt\hbox{$\scriptstyle #2$}}
   {\raise.4pt\hbox{$\scriptscriptstyle #1$}/%
     \lower.4pt\hbox{$\scriptscriptstyle #2$}}}}

\title{On the Geometry of the Berry-Robbins Approach to Spin-Statistics\thanks{A.F. Reyes-Lega gratefully acknowledges
financial support from Universidad de los Andes and from Colciencias-DAAD, through Procol contract no.
373-2007.}}
\subtitle{Spin Stat 2008\\
          Trieste\\
          October 21-25 2008}

\author{Nikolaos Papadopoulos \and    Andr\'es F. Reyes-Lega }
\institute{N. Papadopoulos\at Institut f\"ur Physik THEP, University of Mainz, Germany.\\
\email{papadopoulos@thep.physik.uni-mainz.de}
 \and \\A.F. Reyes-Lega\at
 Departamento de F\'isica, Universidad de los Andes. A.A. 4976, Bogot\'a D.C.\\
  \email{anreyes@uniandes.edu.co}}

\begin{document}
\maketitle
\begin{abstract}
Within a geometric and algebraic framework, the structures which are related to the spin-statistics
connection are discussed. A comparison with the Berry-Robbins approach is made. The underlying geometric
structure constitutes an additional support for this approach. In our work, a geometric approach to quantum
indistinguishability is introduced which allows the treatment of singlevaluedness of wave functions in a
global, model independent way.
\end{abstract}

\section{Introduction}

It is by now   widely accepted that the status of the relation between spin and statistics in
non-relativistic quantum mechanics is quite unsatisfactory for several reasons. Instead of repeating all
these   arguments, we would like to add our motivation for dealing with this problem.   For us, there  are
mainly four reasons:
\begin{itemize}
 \item[$\bullet$] The   influential   approach of Berry-Robbins~\cite{BR}.
 \item[$\bullet$] The $G$-Theory Principle,  as formulated in ~\cite{G1,G2}.
 \item[$\bullet$] We believe that our current understanding of  the spin-statistics connection is not complete
 and that an alternative explanation of it could shed new light into our understanding of quantum
 theory itself.

 \item[$\bullet$]  A re-examination of the spin-statistics connection might be required, in view of new developments
  in theoretical
   physics as, for instance, in the context of higher dimensional theories, quantum gravity, algebraic quantum
   field theory or  non-commutative quantum field theory.
\end{itemize}
The   first two motivations
need  some explanation.

In 1999, one of us (N.P.) had the opportunity to attend a seminar given by Sir Michael Berry in Mainz. At
that time, the impression was that   a   proof of the Spin-Statistics-Theorem within the framework of
non-relativistic quantum mechanics had been established and the intention was to also find a geometric
formulation of this proof. In addition, this seemed to be   a   very good application of the $G$-Theory
Principle. This was formulated in 1987 by a group in Mainz in connection with the reduction of the
Kaluza-Klein Theory in higher dimensions.   The $G$-Theory Principle emphasizes the role of group actions and
symmetries in a maximal way. Therefore, it seemed to be well-suited for the purpose of studying the
spin-statistics connection.  The principle has been applied to the study of anomalies~\cite{G2} and
generalized some years later by J. Sladkowski~\cite{G3}.
  In the present context, one encounters two groups and one manifold: The permutation group  $S_N$ for
$N$ particles, the rotation group $SU(2)$ for spin, and the configuration space $Q$ for $N$ identical
(indistinguishable) particles.
 In this situation, we  first have to clarify the role of the group actions and
manifolds   involved,   before we proceed.

  Regarding the fourth motivation, it comes from previous joint work of one of us (A.R.) with M. Paschke.
In \cite{Paschke}, Paschke proposed to study the spin-statistics connection using the tools of noncommutative
geometry. The idea has been pursued further~\cite{Sp2,RL}, in the hope of eventually establishing a link with
quantum field theory. The idea that this might be possible is based on the fact that the algebraic language
of noncommutative geometry has many features in common with that of quantum field theory~\cite{NONC2}.

  Let us now describe the content of the rest of the paper.   After some preparations in section
\ref{sec_two} in order to fix the notation, we will proceed with the geometric formulation of the problem
  in section \ref{sec_three} and with an equivalent algebraic formulation in section \ref{sec_four}.
  The connection with the Berry-Robbins (BR) approach will be given in section \ref{sec_five}. This
includes a comment on the singlevaluedness condition from the geometric point of view. Finally, in section
\ref{sec_six}, we will present some conclusions.

Lastly, we would like to stress the fact that the entire information of this paper is completely covered by
\mbox{\cite{Sp1,Sp2}}. Therefore, we will be referring   mainly   to the three papers \cite{BR,Sp1,Sp2} and
references   therein, and apologize for not explicitly mentioning a significant number of other interesting
works.

\section{Preparation}\label{sec_two}

We consider $N$ identical particles moving in $ \bb R^3$. The unrestricted configuration space is given by
\begin{align} \widetilde Q_N=\left\lbrace (\vek r_1,\dots,\vek r_N)\in \bb R^{3N}: \vek r_i \neq \vek
r_j\right\rbrace \ .
\end{align}
For $N$
\defemph{identical} particles with the permutation group $G=S_N$ the constrained configuration space is given
by the quotient space   $ Q_N =\xfrac{\widetilde Q_N}{G}$.    In the present discussion we restrict ourselves
to the two particle case, i.e. $N=2$, referring the reader to \cite{RL} for the case of general $N$. For
$N=2$ the effective non-constrained configuration space is given by the sphere $\widetilde Q \equiv
\widetilde Q_2 \cong S^2=\lbrace \vek r \rbrace$.   The exchange of the particles 1 and 2 corresponds to $
\vek r \mapsto - \vek r$ and the underlying permutation group is now given by $G=\bb Z_2$. Therefore, the
constrained configuration space is given by
\begin{align}
 Q\equiv Q_2=\xfrac{\widetilde Q_2}{\bb Z_2}=\xfrac{S^2}{\bb Z_2}= \bb R P^2= \bigl\lbrace[\vek x]\equiv\lbrace\vek x,- \vek x\rbrace: \vek x \in S^2 \bigr\rbrace
\end{align}
 with the injective inclusion $S^2 \inrightarrow \bb R^3 \ , \ \vek r \mapsto \vek
x=(x_1,x_2,x_3)$ . In the standard formalism for two spin $\tfrac{1}{2}$-particles ($s=\tfrac{1}{2}$) the two
spin basis which is fixed is given by
\begin{equation}
\bigl\lbrace | s,m_1\rangle \otimes | s,m_2 \rangle \bigr\rbrace_{m_1,m_2} \quad m_1,m_2 \in \{ \pm
\tfrac{1}{2} \} \mbox{ or, equivalently, by} \bigl\lbrace|j,m\rangle \bigr\rbrace_{j,m}\quad j \in \{0,1\},\
m \in \{-j,\dots,+j\}\ .
\end{equation}
In addition, the wave function is given by
\begin{align}
\Psi(\vek r)=\sum\limits_{m_1,m_2} f_{m_1m_2}(\vek r)\ | s,m_1\rangle \otimes |
s,m_2\rangle=\sum\limits_{j,m} f_{jm}(\vek r) \ | j,m\rangle\ .
\end{align}
The symmetrization postulate corresponds to the choice of the coefficients: The antisymmetric case
$f_{1m}(\vek r)$ and the symmetric case $ f_{00}(\vek r)$ respectively.

In the BR approach, the spin basis is \defemph{moving},
 i.e. $ \bigl\lbrace |s,s;m_1, m_2(\vek r)\rangle  \bigr\rbrace$, or $\bigl\lbrace |j,m(\vek r)\rangle \bigr\rbrace$.
  For this reason, the wave function looks like
\begin{align}
|\Psi(\vek r)\rangle=\sum\limits_{m_1,m_2} \psi_{m_1m_2}(\vek r) \ | s,s;m_1,m_2 (\vek r)
\rangle=\sum\limits_{j,m} \psi_{jm}(\vek r) \ |j,m (\vek r)\rangle,
\end{align}
with the coefficient functions $\psi_{m_1m_2}( \vek r)$ and $\psi_{jm}( \vek r)$. As discussed in BR,
$\psi_{m_1m_2}$ and $f_{m_1m_2}$ are the same functions.

The state $|\Psi\rangle$ lives in a \defemph{two-spin-bundle} over $S^2\,   (\cong \widetilde Q_2)$ along
with the
\defemph{singlevaluedness} constraint given by $|\Psi(\vek r)\rangle=|\Psi(-\vek r)\rangle$.

This is the arena where the BR approach takes place. We denote this by I. We may also think about a second
point of view II, for which the wave function lives in a \defemph{two-spin-bundle}  over $\bb RP^2\,(\cong
Q_2)$ without any constraint. We denote this wave function with $\Phi([\vek x])$. We expect of course an
equivalent situation between the wave functions $|\Psi(\vek r) \rangle \equiv \Psi_I(\vek r)$ and $\Phi([\vek
x])\equiv \Psi_{II}(\vek r)$  and similarly an equivalency for the two vector bundles  $ \eta\equiv\xi_I\cong
S^2 \times V$ respectively $\xi\equiv\xi_{II}=\bb R P^2 \tilde\times V$, where  $V=\bb C^4$ (for
$s=\tfrac{1}{2}$). We expect $\xi$ to be a non-trivial bundle (the symbol $\tilde\times$  will be used to
denote a bundle $M \tilde\times V$  with basis $M$ and fiber $V$ which may or may  not be a trivial bundle).
But what is the precise connection between I and II? This will be clarified in the next section, leading us
to a geometric formulation of the given problem.

\section{Geometric formulation}\label{sec_three}
\subsection{The Two Points of view (I and II)} \label{sub:twopointofview}
As we already saw, two formulations are possible. It is important to realize that
 in the first case (I) the configuration space $Q_I$ is not   constrained, unlike    the associated
 wave function $\Psi_I$,   which is constrained by the singlevaluedness condition.
  In contrast,   in the second case (II) the configuration space $Q_{II}$ is constrained, but
  the wave function    $\Psi_{II}$  is not! For further discussion, it is necessary to keep
  these two points of view in mind. In each formulation, we consider three objects:
\begin{alignat*}{100}
  & && && && \ \, \text{\bf{I}} &&  &&  && \; \text{\bf{II}} \\
  &\text{The basis manifold:} &&\qquad\qquad && \quad Q_I&&\equiv\quad \widetilde Q &&\qquad\qquad && \quad Q_{II}&&\equiv \quad Q\\
  &\text{The two-spin vector bundle: }&& &&\quad \xi_I&&\equiv\quad\eta && && \quad\xi_{II}&&\equiv \quad\xi \\
  &\text{The wave function: }&& &&\Psi_I(\vek r)&&\equiv|\Psi(\vek r)\rangle && &&\Psi_{II}(\vek r)&&\equiv \Phi([\vek x])
\end{alignat*}
In order to find   the equivalence between I and II,   we start from I and construct II. In case I we deal
with an action of the permutation
\defemph{group} $G$. The manifold $\widetilde Q$ is a $G$-space, $G$ is finite and the action of $G$ is free:
\begin{alignat}{100}
&\rho:  &&\quad  && G \times \widetilde Q &&\rightarrow \widetilde Q \ , \quad &&(g,\vek r) &&\mapsto
\rho_g(\vek r)  &&\equiv g \vek r \ \ ,
\end{alignat}
$\eta$ is also a $G$-space, more precisely a $G$-vector bundle.
\begin{dfn}[$G$-Vector bundle]
\label{def:1}
  A   $G$-Vector bundle $(\eta,\tau)$ is given by the following data and   properties:   In an
obvious notation we have $\eta=(E(\eta),\pi, \widetilde Q)$ with fibres $\pi^{-1}(\vek r)=E(\eta)_{\vek
r}\cong V \cong \bb C^n$, the   bundle   projection
\begin{alignat}{100}
&\pi: &&\quad &&E(\eta) &&\rightarrow \widetilde Q \ , \ &&\quad z &&\mapsto \ \pi(z)&&=\vek r \intertext{and
an action $\tau$ of $G$ on the bundle $\eta$} &\tau: && &&G \times \eta &&\rightarrow \eta \ , \ &&(g,z)
&&\mapsto \tau_g(z)&&\equiv gz \ \ .
\end{alignat}
\end{dfn}
\begin{itemize}
\item[(i)] The projection $\pi$ is an equivariant map (consistent with a $G$-action),
 i.e. $\pi \circ \tau_g= \rho_g \circ \pi$.   This is   illustrated by the following commutative diagram:
\begin{center}
    \begin{tikzpicture}
          \matrix (D) [commutative-diagram,row sep=0.8cm]
           {
             |(E1)|   $E(\eta)$ \&
             |(E2)|   $E(\eta)$ \\
             |(tQ1)|  $\widetilde Q$ \&
             |(tQ2)|  $\widetilde Q$ \\
           };
           \begin{scope}[commutative-diagram-arrows]
             \draw[->]         (E1)     --  node[above] {$\tau_g$}   (E2);
             \draw[->] (E1)     --  node[left] {$\pi$}         (tQ1);
             \draw[->] (E2)     --  node[right] {$\pi$}         (tQ2);
             \draw[->]         (tQ1)  --  node[above]  {$\rho_g$} (tQ2);
           \end{scope}
    \end{tikzpicture}
\end{center}
\item [(ii)] The map $\tau_g: \ E(\eta)_{\vek r} \rightarrow E(\eta)_{g\vek r}$ is  a vector space isomorphism.
\end{itemize}
An equivalence (isomorphism) of two $G$-bundles $(\eta,\tau)$ and $(\eta',\tau')$ is a vector bundle
isomorphism $\phi:\eta\rightarrow \eta'$ that commutes with the two actions $\tau$ and $\tau'$, i.e., such
that $\tau'_g\circ\phi(z)=\phi\circ\tau_g(z)$. In this category, we denote a $G$-bundle isomorphism by $\eta
\cong_G \eta'$. The wave function $\Psi_I$ is an element of the space of sections $\Gamma(\eta)$ in $\eta$,
i.e. $\Psi_I \in \Gamma(\eta)$. As explained below,   it turns out that in our case $\Psi_I$ is a
$G$-invariant section in $\eta$: $\Psi_I \in \Gamma^\text{inv}(\eta) $.

At this stage, it becomes evident that we may obtain case II from I by a \defemph{quotienting} procedure.
 Consequently we have:
\begin{alignat*}{100}
&Q_{II}&&=\xfrac{Q_I}{G}&&\quad&&\text{with the projection}&&\quad&& q: \quad && Q_I \rightarrow \xfrac{Q_I}{G}\quad &&(=Q_{II}) \qquad \text{and}\\
&\xi_{II}&&=\xfrac{\xi_I}{G}&& &&\text{with the projection} && && \overline q: && \xi_I \rightarrow \xfrac{\xi_I}{G} &&(=\xi_{II}) \ \ .
\end{alignat*}
The wave function $\Psi_{II}([\,\cdot\,])\in \Gamma(\xi_{II})$ is now completely unconstrained. From the
knowledge of $\xi_{II}$ we can take its pull-back and obtain $\xi_{I}$ from it, as follows. The bundle
$q^*\xi_{II}$ will inherit, in a natural way, a structure of $G$-bundle. We will call $\tilde\tau$ the
corresponding action. It is precisely this $G$-bundle structure that lies behind the constraint condition for
the wave function $\Psi_{I}$.
\begin{rmk} It is important to realize that the $G$-bundle structure induced on $q^*\xi_{II}$ depends on the
topology of the bundle $\xi_{II}$: Two non-isomorphic bundles $\xi_{II}$ and $\xi_{II}'$ will necessarily
give rise to non-isomorphic $G$-bundles $(q^*\xi_{II},\tilde\tau)$ and $(q^*\xi_{II}',\tilde\tau')$,
\emph{even} if the bundles $q^*\xi_{II}$ and $q^*\xi_{II}'$ happen to be isomorphic bundles. The natural
action induced by the pull-back is given by
\begin{equation}\label{eq:tau}
\tilde\tau_g(\vek x,z):=(\rho_g(\vek x),z),
\end{equation}
where $(\vek x,z)\in E(q^*\xi_{II})\subset Q_I\times \xi_{II}$.
\end{rmk}

In order to simplify, we introduce the notation: $Q:=Q_{II}\ , \ \xi:=\xi_{II}\ , \ \xi=(E(\xi),\pi,Q)$.
 Notice that as a result of the quotient operation, the group $G$ is \defemph{not} directly acting on $Q$ and $\xi$
 (the original action of $G$ on $Q_I$ will, nevertheless, be related to the holonomy of the bundle $\xi$).
  The precise connection between I and II is given by   the following
  well-known theorem.
\begin{thm}[cf.~\cite{K}]\label{thm:atiyah}
 If $G$ acts freely on $Q_I$ then there is a \defemph{bijective correspondence} between $G$-bundles ($\xi_I$) over $Q_I$ and bundles
$\xi_{II}=\xfrac{\xi_I}{G}$\  over\ $Q_{II}=\xfrac{Q_I}{G}$\ .
\end{thm}
  Therefore, we can finally express the correspondence between I and II by means of quotient and
pull-back operations, i.e.   $\xi_{II}\cong\xi_I/G$ and $\xi_I\cong_G q^* (\xi_{II})$.

If we start with case I we have:
\begin{alignat}{100}
& \xi_I &&\ \underset{\overline q}{\longrightarrow}\  &&\xfrac{\xi_I}{G} \  &&(=\xi_{II} ) \quad && \text{and} && \quad Q_I && \ \underset{q}{\longrightarrow}\ && \xfrac{Q_I}{G} &&\ (=Q_{II})
\end{alignat}
So we construct the bundle $\xi_{II}$ by taking the quotient with respect to the original $G$-action $\tau$
on $\xi_I$. Again $\xi_I$ can be obtained by a pull-back: $(\xi_I,\tau) \cong_G (q^* (\xi_{II}),\tilde
\tau)$\ .

If we start with case II, i.e. $\xi_{II}=(E(\xi_{II}),\pi,Q_{II})$ we obtain the $G$-bundle $(q^*
(\xi_{II}),\tilde \tau)$ by a pull-back. Furthermore by taking the quotient (\emph{with respect to
$\tilde\tau$}) we achieve again $ \xfrac{q^*( \xi_{II})}{G}\cong \xi_{II} $. This pull-back construction is
expressed by the following diagrams:
  \begin{center}
  \begin{equation}
    \begin{tikzpicture}
          \matrix (D) [commutative-diagram,row sep=0.1cm]
           {
             |(none1)|   $ $ \&
             |(xiII)|    $\xi_{II}$ \&
             |(qxiII)|   $q^*\xi_{II}$ \&
             |(xiII2)|   $\xi_{II}$ \\
         |(fill0)| $ $ \&
         |(fill1)| $ $ \&
             |(fill2)| $ $ \&
             |(fill3)| $ $ \\
             |(Q1)|   $Q_I$ \&
             |(Q2)|  $Q_{II}$ \&
             |(Q12)|  $Q_I$ \&
             |(Q22)|  $Q_{II}$ \\
             |(Q1x)|  $\vek x$   \&
             |(Q2q)|  $q(\vek x)=[\vek x]$ \&
             |(none2)| $ $ \&
             |(none3)| $\hspace{-55px} E(q^*(\xi_{II}))_{\vek x}:=E(\xi_{II})_{[\vek x]}$ \\
           };
           \begin{scope}[commutative-diagram-arrows]
             \draw[->] (xiII)     --  node[right] {$\pi$}  (Q2);
             \draw[->] (qxiII)     --  node[left] {$\pi$}  (Q12);
             \draw[->] (xiII2)     --  node[right] {$\pi$} (Q22);
             \draw[->]         (Q1)    -- node[below] {$q$}       (Q2);
             \draw[->]         (Q12)    -- node[below] {$q$}       (Q22);
             \draw[|->]        (Q1x)   -- node[above]  {$ $}       (Q2q);

           \end{scope}
    \end{tikzpicture}
    \end{equation}
 \end{center}
\begin{rmk}

At this point it may be useful to enumerate all the actions of the permutation group $G$ we use. $G$ acts on
$\widetilde Q\ (\equiv Q_I),\ \eta\ (\equiv\xi_I),\ C(\widetilde Q)$
 -the space of continuous functions on $\widetilde Q$- and $ \Gamma(\eta) $ -the space
  of sections in $\eta$-.   Taking into account that in our case the bundle $\eta$ is a
  trivial bundle (since we are dealing with flat bundles and $\widetilde Q$ is simply-connected),
   we have, in an obvious notation:
\begin{alignat*}{100}
&\rho:&&\quad&& \quad G \times \widetilde Q && \rightarrow &&\ \ \ \widetilde Q\\
&\tau:&&     && \quad G \times \eta     && \rightarrow &&\ \ \ \eta   && \qquad &&(g,z)   && \mapsto && \ \
\tau_g(z)&&:=\quad \tau_g(\vek x,\vek y)
&&\equiv(\rho_g \vek x, R(\vek x,g) \vek y) \qquad \bigl(\text{with }  z=(\vek x,\vek y)\bigr)\\
&     &&     && G \times C(\widetilde Q)    && \rightarrow && C(\widetilde Q) && && (g,a)  &&
\mapsto && (ga)(\vek x)&&:=\ \ \ a\bigl(\rho_{g^{-1}}(\vek x)\bigr) &&\equiv a(g^{-1}\vek x) \\
&     &&     && G \times \Gamma(\eta)   && \rightarrow && \Gamma(\eta) && &&
 (g,s) && \mapsto  && (\hat gs)(\vek x)&&:=\tau_g\bigl(s(g^{-1}\vek x)\bigr)&&\equiv\tau_g\bigl(g^{-1}
  \vek x, |s(g^{-1} \vek x)\rangle\bigr).
\end{alignat*}
  Furthermore, as will become apparent below,   the section $s\ (s(\vek x)\equiv(\vek x,|s(\vek
x)\rangle)$ in $\eta$ is invariant: $ \hat gs=s\ (s\in \Gamma^{\text{inv}}(\eta)) $ if $ |s(g\vek
x)\rangle=R(\vek x,g)|s(\vek x)\rangle$ is valid.
\end{rmk}
\begin{rmk} Since in our case the bundle $\eta$ is trivial, it is legitimate to express elements $z\in\eta$
(globally) in the form $z=(\vek x,\vek y)$. It then follows from Definition \ref{def:1} that  $\tau_g$ must
take the form given above, i.e. $\tau_g(\vek x,\vek y) =(\rho_g \vek x, R(\vek x,g) \vek y)$, with $R$ a
linear mapping -that in general depends on \emph{both} $\vek x$ \emph{and} $g$- taking the fibre over $\vek
x$ onto the fibre over $\rho_g(\vek x)$.
\end{rmk}

\subsection{The Spin zero case as example}\label{sub:spinzeroex}

We consider two spin $S=0$ particles. The permutation group is $G=\bb Z_2$.
We start with the point of view II. For scalar particles, we expect line bundles ($V=\bb C$) over
 the projective space $\bb RP^2$ . It is a   well known    fact that
 there are two line bundles over $\bb  R P^2$ \cite{Husem}, a trivial $\xi_+=\bb RP^2 \times V$ and a
  non-trivial one: $\xi_-=\bb RP^2\tilde\times V$. This corresponds, as we shall see, to symmetric
  and antisymmetric functions on the sphere $S^2$. In this way, we obtain as a direct consequence of
   the non-trivial topology of the configuration space $Q\ (=Q_{II})$ the    Bose-Fermi   alternative for scalar particles, a well known result \cite{FFI,TIP} .

From the point of view I, we now have two  $G$-line bundles:  $\eta_+=(\eta,\tau_+)$ and
$\eta_-=(\eta,\tau_-)$, both with the underlying trivial bundle $\eta = S^2 \times V$. The permutation group
$G=\bb Z_2$ acts non-trivially only on the second line bundle. We denote this action with $\tau_-$.
Explicitly, we have:
\begin{eqnarray}
\tau_-: \;\;\bb Z_2 \times \eta & \longrightarrow & \;\;\;\;\; \eta \nonumber\\
(g,(\vek x,\vek y)) & \longmapsto & (\rho_g(\vek x),\mbox{sign}(g)\vek y).
\end{eqnarray}
 The action $\tau_+$ is trivial. An explicit construction of the non-trivial bundle
$\xi_-$ follows. This is also needed in the next section.

For the projective space $\bb R P^2=\xfrac{S^2}{G}=\left\lbrace [\vek x] \right\rbrace$ we choose the three local charts $(U_\alpha,h_\alpha) , \ \alpha \in \{1,2,3 \}$ as follows:
\begin{align}
 S^2=\bigl\lbrace \vek x=(x_1,x_2,x_3) : x_1^2+x_2^2+x_3^2=1\bigr\rbrace \subset \bb R^3
\end{align}
with the canonical projection $ q: \  S^2 \rightarrow \bb RP^2 \ , \ \vek x \mapsto [\vek x]=\lbrace\vek
x,-\vek x\rbrace=\bb R \vek x$ and the following open covering  of $\bb RP^2$:\\ \mbox{$U_\alpha=\{[\vek x] :
x_\alpha\neq 0\}$}. For example, we may take a look at the first chart, i.e. $\alpha=1$:
\begin{align}
h_1:\ U_1 \rightarrow \bb R^2 \ , \  [\vek x] \mapsto \left(\frac{x_2}{x_1},\frac{x_3}{x_1} \right)
\end{align}
It is convenient to construct the bundle $\xi_-=(E(\xi_-),\pi,\bb RP^2)$ as a non trivial sub-bundle of a
higher rank  trivial bundle $\bb RP^2 \times \bb C^k$. In the construction that follows, we will choose
$k=3$. Consider now a function
\begin{equation}
|\chi (\cdot)\rangle: S^2\rightarrow \bb C^3,
\end{equation}
with the following properties:
\begin{itemize}
\item[(i)] $\langle\chi(\vek x)|\chi(\vek x)\rangle =1$ for all $\vek x$ in $S^2$.\\
\item[(ii)] $|\chi(-\vek x)\rangle= -|\chi(\vek x)\rangle$ for all $\vek x$ in $S^2$.
\end{itemize}
Two possible choices for such a function are $|\chi(\vek x)\rangle= \vek x$ and $|\chi(\vek
x)\rangle=(e^{-i\varphi}\frac{\sin\theta}{\sqrt{2}},-\cos\theta,-e^{i\varphi}\frac{\sin\theta}{\sqrt{2}})$.
For the explicit computations that follow, we will stick to the first choice. Define now the total space of
the bundle as the subset of $\bb RP^2\times \bb C^3$ given by
\begin{align}
 E(\xi_-)=\bigl\lbrace \left([\vek x],\ \lambda | \chi(\vek x) \rangle \right) \in \bb R P^2 \times  \bb C^3:\ \lambda \in \bb C ,\ \vek x \in [\vek
 x]\bigr\rbrace.
\end{align}
Notice that, because of properties (i) and (ii), the fibre over $[\vek x]$ is the complex line in $\bb C^3$
generated by the vector $|\chi(\vek x)\rangle$, \emph{independently} of the choice of representative $\vek
x\in[\vek x]$. However, one must be aware of the fact that, in order to explicitly exhibit an element $z\in
E(\xi_-)$, a choice of representative must be made. If we make the choice $|\chi(\vek x)\rangle$, then (given
$z$) there is exactly one $\lambda\in\bb C$ such that  $z=([\vek x],\ \lambda | \chi(\vek x))\rangle$. On the
other hand, if we  choose to express $z$ in terms of $|\chi(-\vek x)\rangle$, we will find a unique
$\lambda'\in \bb C$ such that $z=([\vek x],\ \lambda' |\chi(-\vek x)\rangle)$. From the definition of
$E(\xi_-)$ and the properties of $|\chi\rangle$, it follows that $\lambda'=-\lambda$. Therefore, we can
indistinctly write $z=([\vek x],\ \lambda | \chi(\vek x))\rangle)=([\vek x],\ -\lambda | \chi(-\vek
x))\rangle)$. The bundle projection is of course defined by $\pi(z)=[\vek x]$. A description of this bundle
in terms of transition functions is now easy to obtain. For this we define  the following local
trivializations:
\begin{align}
 \phi_\alpha: \ \pi^{-1}(U_\alpha) \rightarrow U_\alpha \times \bb C \ , \ ([\vek x],\lambda |\chi(\vek x)\rangle) \mapsto ([\vek x], \mathrm{sign}(x_\alpha)\lambda)\equiv([\vek x], v)
\end{align}
and the   corresponding    transition functions:
\begin{align}
 \phi_\beta \circ \phi_\alpha^{-1}:\ ([\vek x],v)  \mapsto ([\vek x],g_{\beta\alpha} v) \quad\text{with}\quad
g_{\beta \alpha}:\ U_\alpha \cap U_\beta \rightarrow \bb C^\times \ , \  [\vek x] \mapsto
\mathrm{sign}(x_\alpha  x_\beta).
\end{align}
It will be convenient, for the discussion that follows, to have an explicit description, in this geometric
setting, of the space of sections of the bundle just described. First define, for
$\alpha\in\lbrace1,2,3\rbrace$:
\begin{align}
\label{eq:e_i} e_\alpha([\vec x]):= \left(\begin{array}{c} x_\alpha x_1 \\x_\alpha x_2 \\ x_\alpha x_3
\end{array}\right).
\end{align}
These maps can be used to define local sections:
\begin{eqnarray}
\label{eq:s_i}
s^{\mbox{\tiny loc}}_\alpha: U_\alpha &\longrightarrow & \;\;\;\;\;U_\alpha\times \bb C^3\nonumber\\
\left[\vek x\right] &\longmapsto & s_\alpha([\vec x]):=([\vek x], e_\alpha([\vek x])).
\end{eqnarray}
Each $s^{\mbox{\tiny loc}}_\alpha$ can be smoothly extended to a global section $s_\alpha\in \Gamma(\xi_-)$.
Observe that $s_\alpha$ is non-vanishing inside $U_\alpha$, but vanishes exactly outside it, reflecting the
fact that $\xi_-$ is not a trivial (line) bundle. We therefore see that the three sections $s_1,s_2$ and
$s_3$, act as generators of the space of all sections, i.e., every global section $s\in\Gamma(\xi_-)$ can be
written in the form
\begin{align}
s=\sum_{\alpha=1}^3 f_\alpha s_\alpha,
\end{align}
with $f_\alpha\in C(\bb RP^2)$.

Let us now consider the pull-back bundle $q^*\xi_-$. Its total space is given by the set of all pairs $(\vek
x,z)$ in $S^2\times E(\xi_-)$ such that $q(\vek x)= \pi(z)$. Given a section on $\xi_-$, $s\in
\Gamma(\xi_-)$, we can define the following section on the pull-back bundle $(q^*s\in \Gamma(q^*\xi_-))$:
\begin{align}
(q^*s)(\vek x):=(\vek x, s([\vek x])).
\end{align}
Referring back to (\ref{eq:e_i}) and (\ref{eq:s_i}), we then have
\begin{align}
\label{eq:q*s} q^*s_\alpha(\vek x)=(\vek x, s_\alpha([\vek x]))= (\vek x,(\,[\vek x],e_\alpha([\vek
x])\,))\equiv (\vek x,e_\alpha([\vek x])),
\end{align}
where, in last step, we choose a description of the pull-back bundle as a sub-bundle of a  trivial bundle.
The definition of these ``induced'' sections will be needed in the next section.
\section{Algebraic formulation}\label{sec_four}
\subsection{Geometric - algebraic correspondence}

As our example with spin zero particles in the last section shows, in general it is also possible
 to proceed in our discussion on a general level within the geometric framework.
  We consider in this sense manifolds and vector bundles as geometric objects.
  There is, however, an equivalent algebraic framework with applications in physics, e.g.
   in the non-commutative geometric approach to elementary particle physics.
   To put   it   simply, the algebraic object which corresponds to a  topological space $M$
   (compact and Hausdorff)  as
    is shown by the Gelfand-Naimark theorem, is the algebra
       $C(M)$ of complex continuous   functions on $M$. Similarly,
     for a vector bundle over a   space   $M$, the corresponding algebraic object as shown by the Serre-Swan
      theorem, is a finitely generated projective module over   $C(M)$~\cite{NONC,NONC2}.

A merit of the algebraic formulation is that   point sets are treated in a completely global way, and this
allows a clear structural analysis of the physical problem under consideration.
The price for that is one has to deal    (in the present case) with a projective module over  an algebra of
functions,   an object which is not at all common in physics. For example, a vector bundle is described
analytically by a projector in this formalism.    However,   this is much easier to handle than the usual
framework.   As explained in~\cite{Sp2,RL}, in this formalism,  the equivalence
$\Gamma^{\text{inv}}(\xi_I)\cong \Gamma(\xi_{II})$ between the wave functions living in $\xi_{II}$ and the
$G$-invariant wave functions in $\xi_I$, as was briefly discussed in the previous section, is much more
easily established.

 In the following diagram, we exhibit   the bijective correspondence between the geometric and
algebraic    formulations,   as given by the Serre-Swan theorem.
\begin{table}[!h]
\centering
\begin{normalsize}\begin{tabular}{rlcl}
 & \quad\thinspace   \bf{Geometric formulation} &  & \qquad \qquad\quad\bf{Algebraic formulation}\smallskip\smallskip\\
Objects: & Vector bundle  $\xi=(E(\xi),\pi,M)$.     &$\leftrightarrow $ &
\qquad \ \hspace{-0.6cm}Space $\Gamma(\xi)$ of sections of the bundle $\xi$ over $M$.\smallskip \\
Data: & -  Transition functions $\{g_{\beta\alpha}\}$. &$\leftrightarrow $ & - Algebra of functions on $M$: $A=C(M)$. \\
 & - Partition of unity of $M$ $\{\varphi_\alpha \},$ & & - There is a free $A$-module $\mathcal E$ of the form
 $\mathcal E=A^n$ and \\
 & \ \ with $\sum_{\alpha}|\varphi_\alpha|^2=1$, subordinate & &\ \  a projector
  $p_\xi:\ \mathcal E\rightarrow \mathcal E$, with $\Gamma(\xi)\cong p_\xi(\mathcal E)$. (Here\\
 & \ \ to the cover $\{U_\alpha\}$ of $M$.& & \ \ $n$ is the number of open sets in the covering times\\
 &&& \ \ the rank of the bundle).\\
 &&&  -Projector is given by the $A$-valued block-matrix:\\
 &&& \ \  $(p_\xi)_{\alpha\beta}=|\varphi_\alpha|g_{\alpha\beta}|\varphi_\beta|.$
\end{tabular}
\smallskip\\
 {Serre-Swan Theorem: \it{  Bijective correspondence between geometric and algebraic formulation.}}
\end{normalsize}
\end{table}

   In the algebraic formulation, it is natural to start from the point of view I. Furthermore, it turns
out that the $G$-action on the space of functions $C(Q_I)$ determines all information we may have about the
  points   of view I and II.

To point this out, we put for $\widetilde A:=C(\widetilde Q)$ and $A:=C(Q)$ for the algebra functions in
$\widetilde Q\equiv Q_I$ and $Q\equiv Q_{II}$ correspondingly. As was pointed out in the last section, the
permutation group $G$ is acting on $\widetilde A$ (but not on $A$). So $\widetilde A$ is a representation
space for $G$ and we can consider its decomposition with respect to the unitary irreducible representations
$\mathrm{Irr}(G)$ of $G$. Since $G$ is finite, $\mathrm{Irr}(G)$ is a finite set.

  One of us (A.R.)   showed  that this decomposition also leads to an $A$-module decomposition of
the algebra $\widetilde A$, unique up to module isomorphism ($n_R$ is the dimension of the representation
$R$):
\begin{align}
\widetilde A = \bigoplus\limits_{R \in \mathrm{Irr}(G)} A_R \quad \text{with}\quad
A_R=\bigoplus\limits_{i=1}^{n_R} A_{R,i}
\end{align}
Furthermore,   it was shown   that every $A_{R,i}$ is a finitely generated and projective $A$-module:
\begin{thm}[Reyes-Lega, cf.~\cite{Sp2,RL}]
\label{thm:RL} There is   an integer $N_R$   and a projector $p_R$ so that $A_{R,i}$ can be obtained from
$A^{N_R}$, i.e.,
\begin{align}
A_{R,i}\cong p_R(A^{N_R})
\end{align}
\end{thm}
In other words, this shows   (in a way which, as shown below, is very well-suited for comparison with the BR
approach)   that the   $G$-actions   on $\widetilde Q \ (=Q_I)$ and $\eta \ (=\xi_I)$   give us   all
possible   (flat)   vector bundles which appear in the point of view II:
\begin{align}
\xi \ (=\xi_{II}) \in \bigl\{ \xi_R \bigr\}_{R \in \mathrm{Irr}(G)}.
\end{align}

\subsection{Spin zero case, algebraically}\label{sub:spinzeroalgebra}
Here we essentially consider the algebraic version of subsection \ref{sub:spinzeroex}.
 We use the result of the previous subsection for the permutation group $G=\bb Z_2=\{+1,-1\}$ and
 the algebras $\widetilde A=C(\widetilde Q)\equiv C(S^2)$ and $A=C(Q)\equiv C(\bb R P^2) \equiv A_+$.
 The irreducible representations of $G$ are given by   $\mathrm{Irr}(G)=\{R_+,R_-\}$, where $R_+$ is the trivial
 representation   and $R_-$ is given by $R_-: g \mapsto (-1)^g$.

The module decomposition of the previous subsection now takes a simple form:
\begin{align}
\widetilde A=A_+ \oplus A_- \equiv C_+(S^2) \oplus C_-(S^2) \equiv \text{symmetric} \oplus
\text{antisymmetric}.
\end{align}
$\widetilde A$, $A_+$ and $A_-$ are $A$-modules. The algebraic objects $A_+$ and $A_-$ represent, as we may
infer from the module decomposition and the table in the previous subsection, spaces of sections and
correspond to    the    vector bundles $\xi_+$ and $\xi_-$ (geometric objects). This connection is made
explicit by the projectors $p_+$ and $p_-$. It is of course the non-trivial bundle $\xi_-$ which deserves our
attention. As already mentioned, the full information about the bundle $\xi_-$ is contained in the projector
$p_-$. We shall use the geometric information of $\xi_-$ as given in \ref{sub:spinzeroex} to obtain $p_-$. In
a self-explanatory notation we have with $\alpha,\beta \in \{1,2,3\}$
\begin{align}
\varphi_\alpha([\vek x]):=  \left\lbrace \begin{array}{rl}
\sqrt{x_\alpha^2}, & \mbox{if } [\vek x]\in U_\alpha,\\
0, & \mbox{otherwise} \end{array} \white\right\rbrace \ \ ,\ \ \sum\limits_\alpha
\varphi_\alpha^2=\sum\limits_\alpha x_\alpha^2=\vek x^2=1
\end{align}
and the projector which can be written in components with $\chi: \ \chi(\vek x)=(x_1,x_2,x_3)$:
\begin{align}
\label{eq:p_-} p_-=(p_-)_{\alpha\beta} = g_{\alpha\beta}\varphi_\alpha\varphi_\beta= \mathrm{sign}(x_\alpha
x_\beta) |x_\alpha||x_\beta |=x_\alpha x_\beta = |\chi\rangle\langle\chi | \quad .
\end{align}
Having the projector $p_-$, we obtain the $A$-module    $p_- (A^3)$    from the free module $A^3=\lbrace \vek
f=(f_\alpha): f_\alpha \in A \equiv C(\bb R P^2), \alpha=1,2,3 \rbrace$ by taking those $\vek f \in A^3$
which obey the relation $p_- \vek f= \vek f$:
\begin{align}
  p_-( A^3)   =\bigl\{ \vek f:\ p_- \vek f=\vek f, \ \vek f \in A^3 \bigr\}
\end{align}
The relation between the vector bundle $\xi_-$ and the projector $p_-$ is given by the following isomorphism:
\begin{align}
  p_- (A^3)   \cong \Gamma(\xi_-),
\end{align}
i.e.,  the space of  sections $\Gamma(\xi_-)=\{s\}$ is isomorphic (as a module over $A$) to    $p_- (A^3)$.
Moreover, from theorem \ref{thm:RL} it follows that $p_- (A^3)$ is also isomorphic to the $A$-module
$A_-\equiv C_-(S^2)$. An explicit description of these bijective correspondences follows.
\begin{itemize}
\item \underline{$p_-(A^3)\leftrightarrow\Gamma(\xi_-)$:}\smallskip\\ Recall that any section $s\in \Gamma(\xi_-)$ can be
written in the form $s=\sum_\alpha f_\alpha s_\alpha$, with $f_\alpha\in A=C(\bb RP^2)\cong C_+(S^2)$, and
$s_\alpha$ as defined in (\ref{eq:s_i}). So, if we start with $s$, we obtain three functions $f_1,f_2$ and
$f_3$. Setting $\vek f=(f_1,f_2,f_3)$, one can check, using (\ref{eq:e_i}), (\ref{eq:s_i}) and
(\ref{eq:p_-}), that $p_-\vek f=\vek f$ holds. Hence, the map $s=\sum_\alpha f_\alpha s_\alpha\mapsto \vek f$
gives the bijective correspondence between $\Gamma(\xi_-)$ and $p_-(A^3)$.\bigskip
\item \underline{$A_-\leftrightarrow p_-(A^3)$:}\smallskip\\ Consider now an odd function $a\in
A_-=C_-(S^2)$. Using $x_1^2+x_2^2+x_3^2=1$, we can write $a=\sum_\alpha (x_\alpha a)x_\alpha$. Defining the
even functions $f_\alpha(\vek x):= x_\alpha a(\vek x)$, we see that $a$ can be written as $a(\vek
x)=\sum_\alpha x_\alpha f_\alpha(\vek x)$. Since the  $f_\alpha$ are even, we can regard them as elements of
$A$. Therefore, the bijective map between $A_-$ and $ p_-(A^3)$ is given by $a(\vek x)=\sum_\alpha x_\alpha
f_\alpha (\vek x)\mapsto \vek f=(f_1,f_2,f_3)$.
\end{itemize}
\begin{rmk}
The proof that the module $A_-$ can be interpreted as the space of sections on the non-trivial line bundle
over $\bb R P^2$ was first given by Paschke~\cite{Paschke}, using the $SU(2)$ symmetry of the sphere. In the
present paper, the explicit form of the projector follows from the  proof of theorem \ref{thm:RL}, for which
the permutation group plays a prominent role. The equivalence of the two projectors is explained in
~\cite{Sp2}.
\end{rmk}
We now come to a crucial point: If there is a bijective correspondence between $\xi_-$ and $(q^*\xi_-,\tilde
\tau)$, how does this correspondence look like in the algebraic framework? The answer is obtained from the
following \emph{isomorphism of $C(\bb RP^2)$-modules} (cf.\cite{NONC2}, proposition 2.12):
\begin{eqnarray}\label{eq:NC-pull-back}
\widetilde T: C(S^2)\otimes_{C(\bb RP^2)}\Gamma(\xi_-)&\longrightarrow &\Gamma(q^*\xi_-)\nonumber\\
\sum_{\alpha,k} b_k\otimes s_\alpha \;\;\;\;\;\;\;\;\;\;& \longmapsto & \sum_{\alpha,k}b_k\; q^*s_\alpha.
\end{eqnarray}
In this case, theorem \ref{thm:RL} tells us that $\widetilde A\equiv C(S^2)=A_+\oplus A_-$. From $A\equiv
C(\bb RP^2)\cong A_+$ we then obtain:
\begin{eqnarray}
\Gamma(q^*\xi_-)&\cong&C(S^2)\otimes_{C(\bb RP^2)}\Gamma(\xi_-) \cong \widetilde A\otimes_A \Gamma(\xi_-)=
(A_+\oplus A_-)\otimes_A \Gamma(\xi_-) \nonumber\\
&\cong& \left(A_+\otimes_A\Gamma(\xi_-)\right)\oplus\left(A_-\otimes_A\Gamma(\xi_-)\right)\\
 &\cong&
\Gamma(\xi_-)\oplus \left(A_-\otimes_A\Gamma(\xi_-)\right).\nonumber
\end{eqnarray}
This means that it is possible to find an isomorphic copy of $\Gamma(\xi_-)$ \emph{inside} $\Gamma(q^*\xi_-)$
or, in simpler words, every section of $\xi_-$, which is a bundle over $\bb RP^2$, can be expressed as a
certain section on a bundle (the pull-back of $\xi_-$) over $S^2$. All we have to do is to restrict the
domain of $\widetilde T$ to the submodule $\Gamma(\xi_-)$.

 With $T:=\widetilde T\big|_{\Gamma(\xi_-)}$ we then obtain:
\begin{eqnarray}
\label{eq:T}
T: \Gamma(\xi_-) &\longrightarrow& \widetilde T(\Gamma(\xi_-))\subset \Gamma(q^*\xi_-)\nonumber\\
 s=\sum_\alpha f_\alpha s_\alpha & \longmapsto & T(s)=\sum_\alpha f_\alpha\; q^*s_\alpha.
\end{eqnarray}
  Since
sections on $\xi_-$ are unconstrained, we expect to be able to find the correct constraint condition on an
arbitrary section of $q^*\xi_-$, in order to be able to regard it as a section on $\xi_-$. From
(\ref{eq:NC-pull-back}) and (\ref{eq:T}) it is clear that the constraint is the following:
 A section $\sum_\alpha b_\alpha q^*s_\alpha\in \Gamma(q^*\xi_-)$ is the isomorphic image of a section
 in $\Gamma(\xi_)$ if and only if the $b_\alpha$ are \emph{even} functions.This can be recast in terms of the
induced $G$-action $\tilde \tau$ (cf.\cite{Sp2,RL}):
\begin{align} \mbox{A section $\sigma\in \Gamma(q^*\xi_-)$
belongs to the image of $T$ if, and only if, it is $G$-invariant: }\;\;\;\;\; \hat g \sigma = \sigma.
\end{align}
\subsection{Results}
With the above information, it is not difficult to obtain the following results which were derived and
 discussed in more detail in \cite{Sp2,Paschke} (see also \cite{RL}, for the general case).
\begin{itemize}
\item The isomorphism $A_- \cong p_-(A^3) \cong \Gamma(\xi_-)$:\\
As already seen, the space of antisymmetric functions on the sphere $A_-=C_-(S^2)$ can also be described with
the projector $p_-$ and the vector-space-like space $A^3$   (i.e. free module )   by means of the projector
$p_-(A^3)$.     As a consequence, we have the isomorphisms:
\begin{align}
A_-\cong p_-(A^3)\cong \Gamma(\xi_-)\cong T(\Gamma(\xi_-))\;(\subset \Gamma(q^*\xi_-)).
\end{align}
These isomorphisms can be described with the help of the corresponding generators.
 In the case of the above $A$-modules, we do not have a basis at our disposal. Hence, we obtain for the generators and the elements the following expressions:
\begin{alignat}{100}\label{eq:bijections}
&\text{$A$-module:} &&\quad && \qquad\;\; A_- &&\ \longleftrightarrow\ &&\quad p_-(A^3) &&\
\longleftrightarrow\ &&
\quad\Gamma(\xi_-) && \longleftrightarrow\ &&\quad T(\Gamma(\xi_-))\nonumber\\
&\text{Generators:} &&\quad && \qquad\{x_\alpha \}_\alpha &&\ \longleftrightarrow\ &&\{ x_\alpha |\chi (\vek
x)\rangle \}_\alpha
&&\ \longleftrightarrow\ && \quad\{s_\alpha\}_\alpha && \longleftrightarrow\ && \quad\{q^*s_\alpha\}_\alpha\\
&\text{Elements:} &&  && a=\sum_\alpha(x_\alpha a)x_\alpha &&\ \longleftrightarrow\ &&
 \vek f=(f_1,f_2,f_3) && \ \longleftrightarrow\ && s=\sum\limits_{\alpha} f_\alpha s_\alpha && \longleftrightarrow\ &&
 T(s)=\sum\limits_{\alpha} f_\alpha \;q^*s_\alpha,&&\nonumber
\end{alignat}
with $a\in A_-, f_\alpha\in A_+\cong A, f_\alpha(\vek x)=x_\alpha a(\vek x)\leftrightarrow a(\vek
x)=\sum_\alpha x_\alpha f_\alpha (\vek x)$ and $p_-\vek f= \vek f$.
\item The connection in $\xi_-$:\\
A natural connection $\nabla$ in $\xi_-$ is the Grassmann connection which we can also express with the help of the projector $p_-$. So we have for a section $s \in \Gamma(\xi_-)$ as denoted above the relation $ \nabla s \leftrightarrow p_- \mathrm d \vek f \ \ (\nabla^{\xi_-}=\nabla^{p_- A^3})$. This connection is flat and its holonomy group is $\bb Z_2$. \smallskip
\item The bundle $\xi_-$ is a $SU(2)$ bundle:\\
It can be shown that the group $SU(2)$ is acting on $\xi_-$ and we have a $SU(2)$-bundle
 in the sense of the definition in  subsection \ref{sub:twopointofview}.
 Parallel transport by means of the above connection $\nabla$ is consistent with the $SU(2)$ action.
  This point is important for the exchange mechanism of Berry-Robbins.
    Similar conditions were    demanded and discussed in some detail in \cite{BR}.
\end{itemize}

\section{Connection with the Berry-Robbins approach}\label{sec_five}
The aim of this section is to discuss the geometric structure which is behind the Berry-Robbins approach. In
our opinion, this strengthens the relevance of the Berry-Robbins approach to the spin-statistics problem. For
our discussion we need some preparation, we therefore start first with a short review of this approach. In
subsection \ref{sub:constoftwospinbundle}, we give an explicit construction of the \emph{two-spin bundle}
over the projective space $\bb RP^2$ which corresponds to the point of view II of   subsection
\ref{sub:spinzeroex}. In subsection \ref{sub:singlevaluecond} we comment on the singlevaluedness condition,
which   plays    a central role
 in the Berry-Robbins approach, from the geometric point of view.

\subsection{Short review of the exchange mechanism in the Berry-Robbins approach}

We consider two spin $s$  particles. As we already discussed in section \ref{sec_two}, the BR approach refers
to the point of view I, which means that the wave function is essentially defined on the two sphere
$Q_I\equiv \widetilde Q = S^2$ or equivalently on the corresponding trivial vector bundle
$\xi_I\equiv\eta=S^2 \times V$.

The standard spin basis (fixed) is given by:
\begin{align}
| sm_1\rangle \otimes |sm_2\rangle \equiv |m_1 m_2\rangle =:|M\rangle
\end{align}
The permutation of the particles 1 and 2 ($(1,2) \mapsto(2,1)$) leads to
\begin{align}
| M \rangle\equiv | m_1 m_2 \rangle \mapsto |m_2 m_1\rangle =:| \overline{M} \rangle
\end{align}
In order to perform the permutation in a continuous way, an exchange group $G'=SU(2)$ was introduced by BR.
The exchange rotation is then represented by U. In the parametrization of BR this is given by the map
\begin{align}
U: \ S^2 \rightarrow GL(V) \ , \ \vek r \mapsto U(\vek r)= \exp(-\theta\, \vek n(\vek r)\cdot \vek E)
\quad\text{with}\quad \vek n = \vek e_3 \times \vek r\ ,\ \vek r=\vek r(\theta,\varphi). \label{lab:onestar},
\end{align}
where $\vek E$ is a vector operator constructed from the Schwinger representation of spin (cf.\cite{BR}).
With $U(\vek r)$ defined this way, the transported spin basis $|M(\vek r)\rangle $ was defined by
\begin{align}
|M(\vek r)\rangle :&= U(\vek r) | M \rangle \label{lab:twostar}
\intertext{From equation (\ref{lab:onestar}) and (\ref{lab:twostar}) the relation}
| \overline{M} (-\vek r) \rangle &=(-)^{2s} | M (\vek r) \rangle \label{lab:threestar}
\end{align}
for the transported spin basis was obtained. The properties of the transported spin basis   of BR can be
summarized as follows:
\begin{itemize}
\item The smooth map for all $M$:
\begin{align*}
\ S^2 \rightarrow \bb C^{N_s}\ , \ \vek r \mapsto | M(\vek r)\rangle:=U(\vek r)| M \rangle\
\end{align*}
\item The following \emph{exchange} rule:
\begin{align*}
\ | \overline M (-\vek r) \rangle = (-)^{2s} | M(\vek r) \rangle
\end{align*}
\item The \emph{parallel transport} condition:
\begin{align*}
\ \langle M'(\vek r(t)) | \frac{\mathrm d}{\mathrm dt} M (\vek r(t))\rangle=0 \
\end{align*}
 for all $M$ and $M'$, and for every smooth curve $ t \mapsto \vek r(t)$.
\end{itemize}
The wave function is given by
\begin{align}
| \Psi(\vek r) \rangle = \sum \limits_M \psi_M ( \vek r) | M (\vek r) \rangle
\end{align}
In addition, since we have here the point of view I,  the following singlevaluedness condition is imposed in
order to incorporate the indistinguishability of the particles in the formalism:
\begin{align}
|\Psi (- \vek r)\rangle = | \Psi(\vek r) \rangle \label{lab:fourstar}
\end{align}
Assuming the above properties, a direct consequence of equation (\ref{lab:fourstar})  is the relation
\begin{align}
\psi_{\overline M} (-\vek r) = (-)^{2s} \psi_M(\vek r) \label{lab:fivestar}.
\end{align}
This is the correct relation between spin and statistics.

\subsection{Construction of the two-spin bundle} \label{sub:constoftwospinbundle}
The relevance of the \emph{two-spin bundle} over the projective space $\bb R P^2$ was pointed out in BR.
 Here we give an explicit construction of    it for $s=1/2$, using    the geometric and algebraic    formulations
  discussed in the previous sections.
  This construction allows the clarification of the geometric structure which is behind the BR approach.
  In particular, the relation of the exchange mechanism as given by the exchange rotation $U(\vek r)$ to
  topology and geometry (connection and parallel transport) of the system will become transparent.
  The exchange matrix $U(\vek r)$, as explained in BR,   acts on    a 10-dimensional space $V$.
  A basis of $V$ in terms of creation and annihilation operators   (Schwinger representation)
  is given in an obvious  notation:

  \begin{eqnarray}
|e_1\rangle :=  a_1^\dagger a_2^\dagger |0\rangle = |+,+\rangle,& \qquad |e_2\rangle :=  b_1^\dagger
b_2^\dagger |0\rangle = |-,-\rangle, & \qquad |e_3\rangle :=  a_1^\dagger b_2^\dagger |0\rangle =
|+,-\rangle,
\nonumber\\
|e_4\rangle :=  a_2^\dagger b_1^\dagger |0\rangle =|-,+\rangle,&  |e_5\rangle := a_1^\dagger b_1^\dagger
|0\rangle,& \qquad|e_6\rangle :=  a_2^\dagger b_2^\dagger |0\rangle,\\
|e_7\rangle :=  \frac{(a_1^\dagger)^2}{\sqrt{2}} |0\rangle, \qquad& |e_8\rangle :=
\frac{(b_1^\dagger)^2}{\sqrt{2}} |0\rangle, &\qquad
|e_9\rangle :=  \frac{(a_2^\dagger)^2}{\sqrt{2}} |0\rangle,\nonumber\\
  &  |e_{10}\rangle:=\frac{(b_2^\dagger)^2}{\sqrt{2}}|0\rangle.\nonumber&
\end{eqnarray}
So we have for the vector space $V$:
\begin{align}
V=\mathrm {span} \bigl( |e_1\rangle,|e_2\rangle, \dots , |e_{10} \rangle\bigr)
\end{align}
The four vectors $|e_1\rangle,|e_2\rangle,|e_3\rangle$ and $| e_4\rangle $ correspond to the usual two-spin basis $|M\rangle=|m_1 m_2\rangle$. The transported spin vectors $|M(\vek r)\rangle$ are maps:
\begin{align}
S^2 \rightarrow V \ , \ \vek r \mapsto | M (\vek r) \rangle
\end{align}
The exchange matrix $U(\vek r)$ significantly simplifies if we use instead of $| m_1 m_2\rangle $ the total spin basis $|jm\rangle$ . For $j=1$ we use the notation $|m\rangle=|jm\rangle $ for $ m \in \{-1,0,+1\}$ and for $j=0$ we take $|00\rangle$. Now we define a new basis of the space $V$: For every fixed $m$ we consider the corresponding exchange triplet:
\begin{align}
B_m:=\left\lbrace |m\rangle^{(-1)}, | m \rangle^{(0)}, | m \rangle^{(+1)}\right\rbrace
\end{align}
The standard (usual) basis vectors $|m\rangle$ are identified by $|m\rangle^{(0)}\equiv |m \rangle$. Hence, we have with $V_m:=\mathrm{span}(B_m)$ an exchange triplet space. There are of course three such subspaces $V_m$ with $m\in\{-1,0,+1\}$. The new basis of V is given in the following basis scheme:
\begin{alignat}{100}
\begin{matrix}[l]
 j=1 &  \left\lbrace \begin{matrix}[l] |-1\rangle \\ |\ 0\ \rangle \\ |+1\rangle \end{matrix} \right. & &\begin{matrix}
  :  \\ : \\ :  \end{matrix} &  & \begin{matrix}[l] |-1\rangle^{(-1)}=|e_8\rangle &,& |-1\rangle^{(0)}=|e_2\rangle & ,
   & |-1\rangle^{(+1)}=|e_{10}\rangle \\
 |\ 0\ \rangle^{(-1)}=|e_5\rangle & , & |\ 0\ \rangle^{(0)}=\frac{1}{\sqrt{2}}(|e_3\rangle+|e_4\rangle) & ,
 & |\ 0\ \rangle^{(+1)}=|e_6\rangle \\ |+1\rangle^{(-1)}=|e_7\rangle & , & |+1\rangle^{(0)}=|e_1\rangle & , & |+1\rangle^{(+1)}=|e_{9 } \rangle  \end{matrix} \\\\
 j=0 &  & &:& & \begin{matrix} \textcolor{white}{|-1\rangle^{(-1)}=|e_8\rangle}
 &\textcolor{white}{,}&|00\rangle=\frac{1}{\sqrt{2}}(|e_3\rangle-|e_4\rangle).
 \end{matrix}
\end{matrix}
\end{alignat}
Note that the second column represents the standard triplet and singlet. In this basis, the matrix $U(\vek r)$ takes a block diagonal form. The restriction of the $U(\vek r)$ to the space $V_m$ is easily obtained by standard procedures \cite{Sp2} so we have for $U(\vek r) \in GL(V_m)$ the matrix:
\begin{align}
U(\vek r)=\begin{pmatrix} \cos^2\tfrac{\theta}{2}&\ \ & - e^{-i\varphi}\frac{\sin \theta }{\sqrt 2} &\ \ & e^{-2i\varphi} \sin^2\tfrac{\theta}{2} \smallskip\\
e^{i\varphi} \frac{\sin\theta}{\sqrt 2} & & \cos\theta & & -e^{i\varphi} \frac{\sin\theta}{\sqrt 2}\smallskip\\
e^{2i\varphi}\sin^2\tfrac{\theta}{2} & & e^{i\varphi} \frac{\sin\theta}{\sqrt 2} & & \cos^2\tfrac{\theta}{ 2}
\end{pmatrix}
\end{align}
and for the transported vectors we obtain:
\begin{alignat}{100}
&|jm(\vek r)\rangle &&=U(\vek r)|jm\rangle &&=-e^{-i\varphi} \frac{\sin\theta}{\sqrt 2}|m\rangle^{(-1)}+\cos\theta|m\rangle^{(0)}+e^{i\varphi} \frac{\sin\theta}{\sqrt 2} |m \rangle^{(+1)}  \label{lab:onesharp}\\
&|00(\vek r)\rangle &&= |00\rangle \nonumber
\end{alignat}
 From this we again immediately obtain for $\theta=\pi$ the exchange rule expressed now for the total spin
basis:
\begin{align*}
j=1: \ |j m (-\vek r) \rangle = - | j m (\vek r) \rangle \quad , \quad  j=0:\ |00 (-\vek r)\rangle=|00(\vek
r)\rangle
\end{align*}
For $j=0$ we had already $|00\rangle=\mbox{constant}.$ It follows also from (\ref{lab:onesharp}) that every
vector $| j m (\vek r)\rangle$ is not vanishing for all $\vek r$ and $m$. Therefore, we can consider the
mapping: $ \vek r \mapsto |j m (\vek r)\rangle$ as a non-vanishing section in the trivial bundle $ S^2 \times
V_m$. This leads to a definition of a line bundle. In this way, we obtain the four line bundles $\eta_{jm}$
as given by
\begin{align}
\eta_{jm}:=\left\{ |jm(\vek r)\rangle \bb C : \  \vek r \in S^2 \right\} \cong S^2 \times \bb C
\end{align}
  This determines the trivial \emph{two-spin bundle} over the sphere $S^2$:
\begin{align}
\eta=\bigoplus\limits_{j,m} \eta_{jm} \cong S^2 \times \bb C^4 \label{lab:twosharp}
\end{align}
\begin{rmk} Since we are assuming point of view I, in order to complete the description of wave functions one
should:
\begin{itemize}
\item[(i)] Indicate the action $\tau_{jm}$ of $G$ on each bundle $\eta_{jm}$. Although this is not explicitly done in the
BR  formalism, it seems natural to assume that the necessary information is ``hidden'' in the exchange
properties of the transported spin basis. We will comment this in more detail in the next subsection.
\item[(ii)] Once the correct action $\tau_{jm}$ has been found, one should regard as \emph{physical wave
functions} only those sections of $\eta$ that are \emph{invariant} with respect to the given $G$-action. This
corresponds to the assertion that the \emph{physical} configuration space is $\bb RP^2$. The way this is done
in the BR formalism is by imposing the singlevaluedness condition $|\Psi(\vek r)\rangle=|\Psi(-\vek
r)\rangle$. Our main concern here will be to interpret this condition in terms of our formalism (the details
are explained in the next subsection).
\end{itemize}
\end{rmk}
We proceed one step further and construct the \emph{two-spin bundle} over the projective space $\bb RP^2$
which corresponds to the point of view II as discussed in   sections   \ref{sec_two} and \ref{sec_three}.
From the above considerations, it is clear that we expect the bundle $\xi$ to be a direct sum of four line
bundles. In analogy to  equation (\ref{lab:twosharp}) we have:
\begin{align}
\xi=\bigoplus\limits_{j,m} \xi_{jm}
\end{align}
What is left is the determination of the line bundles $\xi_{jm}$ for $j=1$. In order to achieve this,
according to the algebraic  formulation in section \ref{sec_four}, we only have to determine the projector
$P(\vek r)$ which corresponds to the bundle $\xi_{jm}$ ($j=1\ ,\ m\in \{-1,0,+1\}$). This information is, as
expected, hidden in the exchange matrix $U(\vek r)$. Within every subspace $V_m$, we consider the projection
onto the space generated by  $|m\rangle=|jm\rangle\equiv|m\rangle^{(0)}$. Its matrix form,  in terms of the
basis $B_m$, is:
\begin{align}
P_0=\left(
\begin{array}{ccc}
0&0&0 \\0&1&0 \\0&0&0
\end{array}
\right).
\end{align}
From this and $U(\vek r)$ we may    define $P(\vek r)$~\cite{Sp2} on each subspace $V_m$ $(j=1)$ :
\begin{align}
P_m(\vek r): = U(\vek r) P_0 U(\vek r)^\dagger\equiv |jm(\vek r)\rangle\langle jm(\vek r)|.
\end{align}
From the explicit construction of the projector $P_m(\vek r)$, we see that the transported spin basis gives
rise to a projector that is exactly the direct sum of three copies of $p_-$, plus a trivial projector
corresponding to the singlet state (this one describes a trivial line bundle over $\bb RP^2$). Therefore we
may write $P=|jm(\vek r)\rangle\langle jm(\vek r)|$ ($j=1$). It is important to note that the components of
$P$ are even functions, so that we can also regard these projectors as describing bundles over $\bb RP^2$:
$P([\vek x])=P_{ij}(x)$ with $P_{ij} \in A\equiv C(\bb RP^2)\equiv C_+ (S^2)$ is valid. Therefore, we have
$P([\vek x])=P(\vek r)$. Its connection to the bundle $\xi_{jm}$ is given by
\begin{align}
P (A^3) \cong \Gamma(\xi_{jm})
\end{align}
From $P (A^3) \cong p_-( A^3)$ we also see that the isomorphism $\xi_{jm} \cong \xi_-$ holds.

Thus, taking into account the results of  subsections \ref{sub:spinzeroex} and \ref{sub:spinzeroalgebra}, the
determination of the two-spin bundle is completed and we have
\begin{align}
\xi \cong \xi_{1\,-1}\oplus\xi_{1\,0}\oplus\xi_{1\,1}\oplus\xi_{0 \, 0}\cong
\xi_-\oplus\xi_-\oplus\xi_-\oplus\xi_+.
\end{align}
All this was achieved  based on the information which was contained in the transported spin basis $|jm(\vek r)\rangle$ using the geometric and algebraic considerations of the previous sections.

As can be explicitly shown, this basis is, in addition, parallel with respect to the Grassmann connection.

\subsection{On the singlevaluedness condition} \label{sub:singlevaluecond}

The singlevaluedness condition $|\Psi(-\vek r)\rangle=|\Psi(\vek r)\rangle$ seems directly evident as it is a
geometric condition. In spite of this it seems also necessary to examine this condition in the light of the
geometric formulation in section \ref{sec_three}, and especially in the light of   theorem \ref{thm:atiyah}.
  For this purpose, a short recapitulation of the results in section \ref{sec_three} will be useful. The
two points of view I and II may now be summarized as follows:
\begin{center}
\begin{tabular}{lcc}
 & \bf{I} & \bf{II} \bigskip\\
Configuration space: & $\widetilde Q\ (=S^2),\ \vek x = \vek r$ & $Q\  (=\bb R P^2),\ [\vek x] $\smallskip\\
Bundle: & $\eta=(E(\eta),\pi,\widetilde Q)$ & $\xi=(E(\xi),\pi,Q)$\smallskip\\
Sections: & $\Gamma^{\text{inv}}(\eta)$ &$ \Gamma(\xi)$ \smallskip\\
Wave function: & $\Psi^{\text{inv}}(\vek x) \equiv \widetilde\Psi(\vek x)=(\vek x,|\Psi(\vek x)\rangle)$&
$\Psi([\vek x])$ (a section in $\xi$)
\smallskip\\
 & with $\hat g \widetilde \Psi= \widetilde \Psi$ & with $\Psi([g\vek x])=\Psi([\vek x])$
\end{tabular}
\end{center}
The important fact is that in the point of view II the wave function $\Psi \in \Gamma(\xi)$ is unconstrained.
The direct \emph{physical} wave function  corresponds in the point of view I to the constrained wave function
\begin{align}
\widetilde \Psi \in \Gamma^{\text{inv}}(\eta) \subset \Gamma(\eta).
\end{align}
As shown above, the condition on $\widetilde\Psi$ is given by the action $\hat g$ of the permutation group
$G$: $ \hat g \widetilde \Psi = \widetilde\Psi $.

The relation between singlevaluedness (as proposed in BR) and invariance of the wave function (as proposed in
the present work) is a very  subtle issue. Therefore, we will spell out this relation in detail, for the case
$s=0$, taking advantage of the results presented in the previous sections. In the next paragraphs, we follow
the notation and conventions of sections 3 and 4.

 Let us start by considering an arbitrary section
$\sigma$ of the pull-back bundle $q^*\xi_-$. We have seen that, in view of (\ref{eq:NC-pull-back}), it can be
written in the form $\sigma=\sum_\alpha b_\alpha\;q^*s_\alpha$, with $b_\alpha\in\widetilde A=C(S^2)$ (here,
the sections $s_\alpha$ denote the generating sections defined in (\ref{eq:s_i})). We have seen that $\sigma$
lies in the image of $T$ if and only if the functions $b_\alpha$ are even. The relation with invariance of
the section is a consequence of the following calculation:
\begin{eqnarray}
\hat g\sigma (\vek x) &=& \tilde \tau_g\sigma(g^{-1}\vek x)= \tilde \tau_g\big(\sum_\alpha
b_\alpha(g^{-1}\vek x)\,q^*s_\alpha(g^{-1}\vek x)\big)\nonumber\\
&=& \tilde \tau_g\big(g^{-1}\vek x,\sum_\alpha b_\alpha(g^{-1}\vek x)e_\alpha(\underbrace{[g^{-1}\vek
x]}_{=[\vek x]})\big)\\
&\stackrel{(\ref{eq:tau})}=& \big(\vek x, \sum_\alpha b_\alpha(g^{-1}\vek x)e_\alpha([\vek x])\big).\nonumber
\end{eqnarray}
We thus see that
\begin{align}
\hat g\sigma= \sigma \quad\Leftrightarrow\quad b_\alpha\in A_+\cong A \quad\Leftrightarrow\quad \sigma\in
T(\Gamma(\xi_-)).
\end{align}
Therefore, if a section $\sigma$ is the image of some $s=\sum_\alpha f_\alpha s_\alpha\in \Gamma(\xi_-)$
($f_\alpha$ must be even), then we can express it as follows:
\begin{eqnarray}\label{eq:sigma}
\sigma(\vek x) &=& \big(\vek x, \sum_\alpha f_\alpha(\vek x)\,q^*s_\alpha (\vek x)\big)\nonumber\\
&=&\big(\vek x, \sum_\alpha f_\alpha(\vek x)e_\alpha([\vek x])\big)\\
&=&\big(\vek x, \sum_\alpha f_\alpha(\vek x)x_\alpha |\chi(\vek x)\rangle\big)\nonumber\\
&=&\big(\vek x, a(\vek x)|\chi(\vek x)\rangle\big).\nonumber
\end{eqnarray}
Here we have made use of the fact that we are working on the pull-back bundle, and in this case it is
possible to express $e_\alpha$ as the product $e_\alpha([\vek x])=x_\alpha|\chi(\vek x)\rangle$ and then to
``absorb'' the term $x_\alpha$ into the function $f_\alpha$, giving place to the \emph{odd} function
$a_\alpha=\sum_\alpha f_\alpha x_\alpha$. This is in full agreement with the bijections described in
(\ref{eq:bijections}). We may conclude: The section $\sigma$ in (\ref{eq:sigma}) is an invariant section (and
hence represents a physical wave function) if and only if each $f_\alpha$ is an \emph{even} function or,
equivalently, if the function $a$ is an \emph{odd} function. The fact that we can factor out this odd
function is due to our choice of $|\chi(\vek x)\rangle$ with the property $|\chi(-\vek x)\rangle=-|\chi(\vek
x)\rangle$. Notice that we are now working on the pull-back bundle, which is a trivial bundle. Whereas being
an invariant section is something independent of the way the pull-back bundle is represented, the fact that
$a$ must be odd in order for $\sigma$ in (\ref{eq:sigma}) to be invariant is something that depends on our
specific construction of the bundle (there are infinitely many bundles that are isomorphic to $q^*\xi_-$, but
``look'' differently).

In order to distinguish the features that depend on a choice from those that do not, we will proceed in the
following way.
\begin{enumerate}
\item Let us assume that the physical wave functions for spin zero particles are sections on the bundle
$\xi_-$ over $\bb R P^2$ (this gives of course the wrong connection between spin and statistics, but the same
exercise could be done with the trivial bundle). In this case we would have, assuming point of view II:
$\Psi_{II}\equiv s \in \Gamma(\xi_-)$.
\item In order to obtain the description of this wave function using point of view I, we take the pull-back
of $\xi_-$ and consider (as we are forced to) the $G$-action $\tilde \tau$ on $q^*\xi_-$ naturally induced by
the pull-back operation (cf. (\ref{eq:tau})). From the definition of pull-back, the  description of
$q^*\xi_-$ as a sub-bundle of a trivial bundle leads naturally to a description of the bundle in terms of the
map $|\chi\rangle$.
\item Study the invariance of the section $\Psi_I\equiv T(s)$ using the $G$-bundle ($q^*\xi,\tilde \tau$)
and compare with the singlevaluedness condition.
\item Construct an isomorphism of $G$-bundles $(q^*\xi_-,\tilde \tau)\cong_G(\eta,\tau')$, with $\eta$
described in terms of a map $|\chi'\rangle$ having the property $|\chi'(\vek x)\rangle= |\chi'(-\vek
x)\rangle$.
\item Study the invariance of the section $\Psi_I\equiv T(s)$ using the bundle ($\eta,\tau'$) and compare with the singlevaluedness
condition.
\end{enumerate}
Let us now go through these five steps:
\begin{enumerate}
\item We start with $\Psi_{II}=s\in \Gamma(\xi_-)$. As explained before, there must be some functions
$f_\alpha\in A\cong A_+$ such that $s=\sum_\alpha f_\alpha s_\alpha$.
\item From (\ref{eq:T}) we obtain $T(s)=\sum_\alpha f_\alpha q^*s_\alpha \in
T(\Gamma(\xi_-))\subset\Gamma(q^*\xi_-)$. From the definition of pull-back, we obtain:
\begin{equation}\label{eq:step1}
E(q^*\xi_-)=\lbrace(\vek x,([\vek x],\lambda|\chi(\vek x)\rangle))\in S^2\times E(\xi_-): \lambda\in \bb
C\rbrace\equiv \lbrace(\vek x,\lambda|\chi(\vek x)\rangle)\in S^2\times E(\xi_-): \lambda\in \bb C\rbrace.
\end{equation}
Using this and (\ref{eq:tau}), we obtain the explicit form of $\tilde \tau$ (for $g=-1$):
\begin{eqnarray}\label{eq:taustep1}
\tilde \tau_g(\vek x,\lambda|\chi(\vek x)\rangle)&=&(\rho_g\vek x,\lambda|\chi(\vek x)\rangle) =(-\vek
x,\lambda|\chi(\vek x)\rangle)\nonumber\\
&=&(-\vek x,-\lambda|\chi(-\vek x)\rangle),\;\;\;\;\;\;\;\;\mbox{i.e., } \tilde\tau\equiv \tau_-.
\end{eqnarray}
\item We must have $\Psi_I=T(s)=T(\Psi_{II})$. Writing $\Psi_I$ as $\Psi_I(\vek x)=(\vek x, |\Psi_I(\vek
x)\rangle)$, we conclude, from (\ref{eq:step1}), that there must be a function $a\in C(S^2)$ such that
\begin{align}
\Psi_I(\vek x)=(\vek x, a(\vek x)|\chi(\vek x)\rangle).
\end{align}
From our previous computations it then follows that $\Psi_I=T(s)$ if and only if $a(\vek x)$ is \emph{odd}.
This in turn implies:
\begin{eqnarray}
|\Psi_I(-\vek x)\rangle & = & a(-\vek x)|\chi(-\vek x)\rangle= \big(-a(\vek x)\big)\big(-|\chi(\vek
x)\rangle\big)\nonumber\\
&=& |\Psi_I(\vek x)\rangle.
\end{eqnarray}
\item From (\ref{eq:step1})  it is clear that if $|\chi'(\vek x)\rangle$ is \emph{any}
non-vanishing, normalized and smoothly varying vector, then replacing $|\chi\rangle$ by $|\chi'\rangle$ in
(\ref{eq:step1}) we obtain a bundle  $\eta$ which is isomorphic to $q^*\xi_-$ (this is a quite obvious fact,
because both bundles are trivial). Now, as we have seen, the $G$-action $\tilde \tau$ on $q^*\xi_-$ is
equivalent to $\tau_-$. Since the equivalence class of this action is completely determined by the (now
fixed) $\xi_-$, the action $\tau'$ must also be equivalent to $\tau_-$. Therefore, we define
\begin{eqnarray}
\tau_g'(\vek x,\lambda |\chi'(\vek x)\rangle):=(g\vek x, \mbox{sign}(g)|\chi'(g\vek x)\rangle).
\end{eqnarray}
It is easy to check   that $(q^*\xi_-,\tilde \tau)\cong_G (\eta,\tau')$. Although  this result is independent
of the specific choice of $|\chi'\rangle$, in the next step we will assume that $|\chi'(-\vek x)\rangle=
|\chi'(\vek x)\rangle$.
\item Again, we must have $\Psi_I(\vek x)=(\vek x,|\Psi_I(\vek x)\rangle)=(\vek x, a(\vek x)|\chi'(\vek x)\rangle)$, for some $a\in C(S^2)$.
We know that $\Psi_I=T(s)$ if and only if $\Psi$ is an invariant section in $(\eta, \tau')$. In this case,
the requirement of invariance leads to ($g=-1$):
\begin{eqnarray}
\hat g \Psi_I(\vek x) &=& \tau_g'(-\vek x, a(-\vek x)|\chi'(-\vek x)\rangle)\nonumber\\
&=& (\vek x, -a(-\vek x)|\chi'(-\vek x)\rangle).
\end{eqnarray}
Hence, $\Psi_I$ is invariant if and only if $-a(-\vek x)=a(\vek x)$, i.e., if and only if $a$ is \emph{odd},
as expected from the bijections in (\ref{eq:bijections}). The unexpected result is the following:
\begin{eqnarray}
|\Psi_I(-\vek x)\rangle & = & a(-\vek x)|\chi'(-\vek x)\rangle= \big(-a(\vek x)\big)|\chi'(\vek
x)\rangle\nonumber\\
&=& -|\Psi_I(\vek x)\rangle.
\end{eqnarray}
\end{enumerate}

The result is that from the above considerations, it is not possible to justify the singlevaluedness
condition. Whereas the singlevaluedness condition can be imposed when the transported vector is chosen to be
$|\chi\rangle$, it cannot be imposed if choose to work with $|\chi'\rangle$. On the other hand, in both cases
the invariance condition leads to a bijective correspondence between the section $s$ and the same odd
function $a$. Our argument can be easily generalized to deal with the general spin case, or with more than
two particles, but we have chosen the spin zero case because of its simplicity and because it already
contains the essential idea.

\section{Discussion}\label{sec_six}
There is no doubt that the understanding of indistinguishability in quantum mechanics is a very subtle problem.
 It is not difficult to accept that the role of indistinguishable particles in the formulation and interpretation
 is very important and lies in the heart of quantum mechanics itself.
 In this sense, it is not understandable why, for instance, quantum field theory, from an outside perspective,
 should explain the spin-statistics connection and not quantum mechanics itself.

In the present contribution we analyzed within a geometric framework (section \ref{sec_three}) and
 in addition within an equivalent algebraic framework (section \ref{sec_four}) the structures related
 to the spin-statistic connection. Although our approach, particularly in sections \ref{sec_three},\ref{sec_four}
 and even subsection \ref{sub:singlevaluecond} on the singlevaluedness condition, is quite general and
 independent of the work of Berry-Robbins, we chose a close reference to it since we find it very
  interesting and inspiring.

In the geometric formulation we point out that there are two points of view when dealing with
indistinguishable particles in quantum mechanics. From the point of view I, which is in essence the usual
point of view, the configuration space $\widetilde Q$ is unconstrained whereas the wave function has to be
constrained. Here e.g. the singlevaluedness condition or another condition may be imposed. From the point of
view II we have the opposite situation: the effective configuration space $Q$ is constrained by
identification as imposed by the permutation group whereas the wave function is now completely unrestricted.
By a wave function in the case of a non-trivial spin bundle we mean a section in a bundle. This dual
situation may cause a lot of confusion. We believe that this was entirely clarified with the help of theorem
\ref{thm:atiyah} and the considerations in section \ref{sec_three}.

We expect that our approach, both geometric and algebraic, will help to clarify in general
 the spin-statistics problem. In particular in section \ref{sec_five}, the connection with the
  Berry-Robbins approach allowed us to clarify the geometric structure and underlines in this sense
  the relevance  of the Berry-Robbins approach.

The concept of the singlevaluedness of the wave function under particle exchange is a subtle one.
 In this work, the geometric approach to quantum indistinguishability allowed us to
 treat the singlevaluedness of the wave function in a global, model independent way.
 The result is that we   cannot   justify this condition from the geometric framework,
 but have to replace it by a less stringent condition:   The   global invariance of the wave
  functions. This does not mean that this condition is wrong.
   From our experience with anomalies \cite{G2} we may expect that there are other physical
   conditions, not known at the moment, which demand and justify the singlevaluedness condition.

\end{document}